\documentclass[aps,prd,reprint,superscriptaddress,nofootinbib]{revtex4-2}
\usepackage{amsmath,amsfonts}
\usepackage{graphicx}
\usepackage[caption=false]{subfig}
\usepackage[dvipsnames,usenames]{xcolor}

\usepackage{makecell}

\usepackage[colorlinks,citecolor=blue]{hyperref}

\newcommand{\blue}[1]{{\color{black}{#1}}}

\begin{document}

\title{Symmetry and bipolar motion in collective neutrino flavor oscillations}

\author{Zewei Xiong}
\email[Email: ]{z.xiong@gsi.de}
\affiliation{GSI Helmholtzzentrum {f\"ur} Schwerionenforschung, Planckstra{\ss}e 1, 64291 Darmstadt, Germany}

\author{Meng-Ru Wu}
\affiliation{Institute of Physics, Academia Sinica, Taipei 11529, Taiwan}
\affiliation{Institute of Astronomy and Astrophysics, Academia Sinica, Taipei 10617, Taiwan}
\affiliation{Physics Division, National Center for Theoretical Sciences, Taipei 10617, Taiwan}

\author{Yong-Zhong Qian}
\affiliation{School of Physics and Astronomy, University of Minnesota, Minneapolis, Minnesota 55455, USA}

\date{\today}

\begin{abstract}
We identify a geometric symmetry on the two-flavor Bloch sphere for collective flavor oscillations of a homogeneous dense neutrino gas.
Based on this symmetry, analytical solutions to the periodic bipolar flavor evolution are derived.
Using numerical calculations, we show that for configurations without this symmetry, the flavor evolution displays deviations from the bipolar flavor motion or even exhibits aperiodic patterns.
We also discuss the implication of our finding for more general three-flavor and inhomogeneous cases.
\end{abstract}

\maketitle
\graphicspath{{./figures/}}

\section{Introduction}
\label{sec:introduction}
Neutrino flavor oscillations in vacuum and ordinary matter are well established by solar, atmospheric, reactor, and accelerator neutrino experiments \cite{pdg}.
In astrophysical environments such as core-collapse supernovae and binary neutron star mergers, where neutrinos are produced copiously, coherent forward scattering couples neutrinos emitted with different energies and directions.
The non-linear nature of this neutrino self-coupling leads to a variety of collective phenomena involving flavor instability and conversion, which can impact the dynamics and nucleosynthesis of those astrophysical environments \cite{duan2011influence,malkus2012neutrino,wu2015effects,sasaki2017possible,wu2017imprints,stapleford2020coupling,xiong2020potential,george2020fast,li2021neutrino,just2022fast,fernandez2022fast,fujimoto2023explosive,ehring2023fast}.

Collective neutrino flavor evolution is in general a highly complicated quantum problem that can be affected by many-body entanglements and correlations \cite{serreau2014neutrino,rrapaj2020exact,patwardhan2021spectral,roggero2021dynamical,roggero2021entanglement,xiong2022many,martin2022classical,roggero2022entanglement,cervia2022collective,lacroix2022role},
non-forward scatterings (collisions) of neutrinos \cite{martin2021fast,hansen2022enhancement,kato2022effects,sasaki2022detailed,padilla2022neutrino2, johns2021collisional,johns2022collisional,lin2022collision,xiong2022collisional},
advection of neutrinos in an inhomogeneous environment \cite{martin2020dynamic,bhattacharyya2021fast,wu2021collective,zaizen2021nonlinear,richers2021neutrino,grohs2022neutrino,richers2022code,abbar2022suppression,shalgar2022supernova,nagakura2022time,xiong2022evolution},
and participation of all three flavors \cite{duan2008stepwise,dasgupta2008collective,friedland2010self,capozzi2020mu,zaizen2021three,shalgar2021three}.
Calculations including all these aspects are practically infeasible at the moment. 
Mean-field approximation, collision-less limit, assumption of homogeneity, and two-flavor simplification were often adopted (although not always all at the same time) in order to elucidate different aspects of the nonlinear system and make the treatment tractable.
In particular, under all the above assumptions, neutrino flavor evolution can be studied in terms of the polarization vectors on a Bloch sphere under the $SU(2)$ algebra \cite{samuel1993neutrino,pastor2002physics}.
We follow this instructive approach in this work.

Depending on the neutrino spectral and angular distributions, various modes of flavor instabilities including the ``slow'' and ``fast'' types have been identified (see e.g., \cite{duan2010collective,tamborra2021new,richers2022fast,capozzi2022neutrino,volpe2023neutrinos} for reviews).
The slow mode requires ``crossings'' in the neutrino energy spectrum while the fast mode demands the same in the neutrino angular distributions \cite{izaguirre2017fast,abbar2018fast,morinaga2022fast,dasgupta2022collective}.
Depending on whether the azimuthal symmetry is spontaneously broken or not, the slow mode can be further subdivided into the multi-zenith-angle (MZA) or multi-azimuthal-angle (MAA) type, while the fast mode can be subdivided into the axially symmetric (AS) or axial-symmetry-breaking (SB) type\footnote{MZA and AS have the same meaning, and so do MAA and SB. Here they are named differently to distinguish the slow from the fast mode.} \cite{duan2006coherent,raffelt2007self,raffelt2013axial,chakraborty2016self,izaguirre2017fast,dasgupta2017fast,yi2019dispersion}.

Flavor evolution induced by some instabilities (e.g., the single-angle slow and single-energy AS fast modes) shows periodic and bipolar behavior, for which in each period neutrinos undergo flavor conversion but soon return to their initial state collectively. 
However, more complicated patterns can arise for other instabilities, e.g., kinematic decoherence in the MZA slow mode \cite{raffelt2007self}, relaxation and cascades in the multi-energy AS fast mode \cite{johns2020fast}, and oscillation around a stationary state in the SB fast mode \cite{xiong2021stationary}.
Those modes either completely lack periodicity or show aperiodic features that mimic periodic behavior only over a short time.

The bipolar motion in the single-angle slow mode was found as early as in Ref.~\cite{pastor2002physics} and can be understood by making an analogy to a gyroscopic pendulum \cite{hannestad2006self,duan2007analysis}.
Several efforts were recently made to probe the underlying physics of the single-energy AS fast mode by counting degrees of freedom or in terms of Gaudin invariants \cite{johns2020neutrino,padilla2022neutrino,fiorillo2023slow}.
In this paper we provide an alternative and more visual picture from the viewpoint of geometric symmetry.

The basic idea is as follows.
Consider periodic motion of neutrino polarization vectors on the surface of the Bloch sphere.
We refer to the geometric shape or distribution formed by the tips of those vectors as a ``configuration'', which can expand or contract collectively during the evolution.
Because those vectors return to their initial state after each oscillation period, one simple way is for them to maintain their initial configuration throughout the evolution.
This configuration can only be a \textit{circular} distribution, which is ensured by the single-angle slow or single-energy AS fast mode.
With this symmetry, we can derive the analytical solution for the periodic bipolar system.
More importantly, this picture helps us understand the distinct behaviors in other aperiodic evolution and makes a connection to more general cases involving three flavors and an inhomogeneous neutrino gas.

In Sec.~\ref{sec:EOM} we present the mean-field equation of motion (EOM) and its linearized stability analysis.
We show the existence of the geometric symmetry in the bipolar flavor evolution in Sec.~\ref{sec:periodic}.
We use numerical examples to demonstrate how the presence (absence) of this symmetry leads to periodic (aperiodic) behaviors of neutrino flavor evolution in Sec.~\ref{sec:aperioic}.
Implications for systems that break the spatial homogeneity and that include three flavors are discussed in Sec.~\ref{sec:conclusions}.

\section{Equation of motion and stability analysis}
\label{sec:EOM}
For a two-flavor system including $\nu_e$, $\nu_x$ ($x=\mu$ or $\tau$) and their antineutrinos, the mean field for neutrinos of momentum $\vec{q}$ can be described by the polarization vector $\mathbf{P}(\omega,\vec{v})$, where $\vec{v}=\vec q/|\vec q|$ is the neutrino velocity, $\omega=\pm \delta m^2/(2|\vec{q}|)$ is the vacuum oscillation frequency, $\delta m^2>0$ is the mass-squared difference between the two mass eigenstates, and the plus (minus) sign is for neutrinos (antineutrinos).
The polarization vector $\mathbf{P}(\omega,\vec{v})$ can be decomposed in the flavor space with three unit basis vectors $\hat{\mathbf{e}}_1$, $\hat{\mathbf{e}}_2$, and $\hat{\mathbf{e}}_3$.  
The vertical component $P_3$ is directly related to the probability of finding a $\nu_e$ or $\nu_x$ while the horizontal component $\mathbf{P}_\perp=P_1\hat{\mathbf{e}}_1+P_2\hat{\mathbf{e}}_2$ measures coherence of flavor evolution.

The time evolution for a collisionless and homogeneous neutrino gas is governed by the EOM
\begin{equation}
	\partial_t \mathbf{P}(\omega,\vec{v}) = \mathbf{H}(\omega,\vec{v}) \times \mathbf{P}(\omega,\vec{v}),
	\label{eq:pvw_vdrho_P}
\end{equation}
where the total Hamiltonian is $\mathbf{H}(\omega,\vec{v}) = \mathbf{H}_\text{vac}(\omega) + \mathbf{H}_\text{mat}(\vec v) + \mathbf{H}_{\nu\nu}(\vec{v})$.
The first term $\mathbf{H}_\text{vac}(\omega)=\omega \mathbf{B}$ accounts for vacuum mixing, where the unit vector $\mathbf{B}$ is $(\sin 2\theta_V,\,0,\,-\cos 2\theta_V)$ for the normal mass ordering and $(-\sin 2\theta_V,\,0,\,\cos 2\theta_V)$ for the inverted mass ordering, with $\theta_V$ being the vacuum mixing angle.
The second term $\mathbf{H}_\text{mat}(\vec v) = v^\rho(\vec{v}) \lambda_\rho \hat{\mathbf{e}}_3$ originates from coherent forward scattering of neutrinos on ordinary matter particles, where $v^\rho(\vec{v})=(1, \vec{v})$ is the four-velocity of neutrinos, $\lambda^\rho=\sqrt{2} G_F n_e v^\rho_\mathrm{bulk}$, $n_e$ is the net electron number density, and $v^\rho_\mathrm{bulk}$ is the four-velocity of matter.
We use the spacetime metric $\mathrm{diag}(+1,\,-1,\,-1,\,-1)$.
Because a large $\mathbf H_\text{mat}$ effectively suppresses $\theta_V$, we assume a very small $\theta_V$ throughout this paper.
The last term $\mathbf{H}_{\nu\nu}(\vec{v})=\mu v^\rho(\vec{v}) \mathbf{J}_\rho$ is due to $\nu$--$\nu$ interaction, where
\begin{equation}
	\mathbf{J}^\rho = \int d\vec{v}\, v^\rho(\vec{v}) \int_{-\infty}^{+\infty} d\omega\, F(\omega,\vec{v}) \mathbf{P}(\omega,\vec{v})
	\label{eq:pvw_Jmu}
\end{equation}
is the neutrino polarization current, $\mu F(\omega,\vec{v})=\sqrt{2}G_F \text{sgn}(\omega)[F_{\nu_e}(\omega,\vec{v})-F_{\nu_x}(\omega,\vec{v})]$, and $F_{\nu_e}(\omega,\vec{v})$ with $\omega>0$ ($\omega<0$) is the $\nu_e$ ($\bar\nu_e$) spectral and angular distribution function. The $\nu_e$ number density is given by $\int d\vec{v}\, \int_0^{+\infty} d\omega\, F_{\nu_e}(\omega,\vec{v})$.
For convenience of presenting numerical examples, we use $\mu^{-1}$ as a typical length scale (in contrast to the usual definition $\mu=\sqrt{2} G_F n_{\nu_e}$ in literature) and note that only the product $\mu F(\omega,\vec v)$ matters.

Defining $\mathsf P_\perp\equiv P_1-i P_2$\footnote{We use the sans serif font to denote the complex functions for the horizontal components in order to distinguish e.g., $\mathsf P_\perp$ from its vector form $\mathbf P_\perp$. The horizontal component $\mathsf P_\perp$ is often written as $S$ in literature. Note that the dependence on $\omega$ and $\vec v$ are often suppressed to save space in this paragraph and Sec.~\ref{sec:periodic}.} and $\mathsf H_\perp\equiv H_1-i H_2$, we rewrite Eq.~\eqref{eq:pvw_vdrho_P} as:
\begin{align}
	i \partial_t \mathsf P_\perp & =
	H_3 \mathsf P_\perp
	- P_3 \mathsf H_\perp, \nonumber\\
	\partial_t P_3 & =
	\mathrm{Im} ( \mathsf H_\perp \mathsf P_\perp^* ).
	\label{eq:pvw_dt_Pperp}
\end{align}
In the limit where $|\mathsf P_\perp| \ll P_3 \approx 1$ for all $\omega$ and $\vec v$, Eq.~\eqref{eq:pvw_dt_Pperp} takes the linearized form
\begin{equation}
	i \partial_t \mathsf P_\perp =
	H_3 \mathsf P_\perp
	- \mathsf H_\perp.
\end{equation}
A stability analysis can be performed using the above linearized EOM and assuming that all neutrinos follow a collective mode with $\mathsf P_\perp = \mathsf Q\, e^{-i\Omega t}$ and $\mathsf H_\perp = \mathsf T\, e^{-i\Omega t}$, where $\mathsf Q$ and $\mathsf T$ are time-independent and $\Omega=\Omega_r+i\Omega_i$.
This procedure gives
\begin{equation}
    \mathsf Q=\frac{\mathsf T}{H_3-\Omega},
    \label{eq:cp_ansatz}
\end{equation}
from which the unstable eigenmode(s) with $\Omega_i>0$ can be found.
For an unstable system, although the initial distribution of $\mathsf P_\perp$ can be different from that of $\mathsf Q$, $\mathsf P_\perp$ quickly take the same shape as $\mathsf Q$ following exponential growth of the unstable mode in the linear regime.

\section{Symmetry in periodic bipolar motion}
\label{sec:periodic}
\begin{figure*}[hbt!]
	\centering
	    \includegraphics[width=0.8\textwidth]{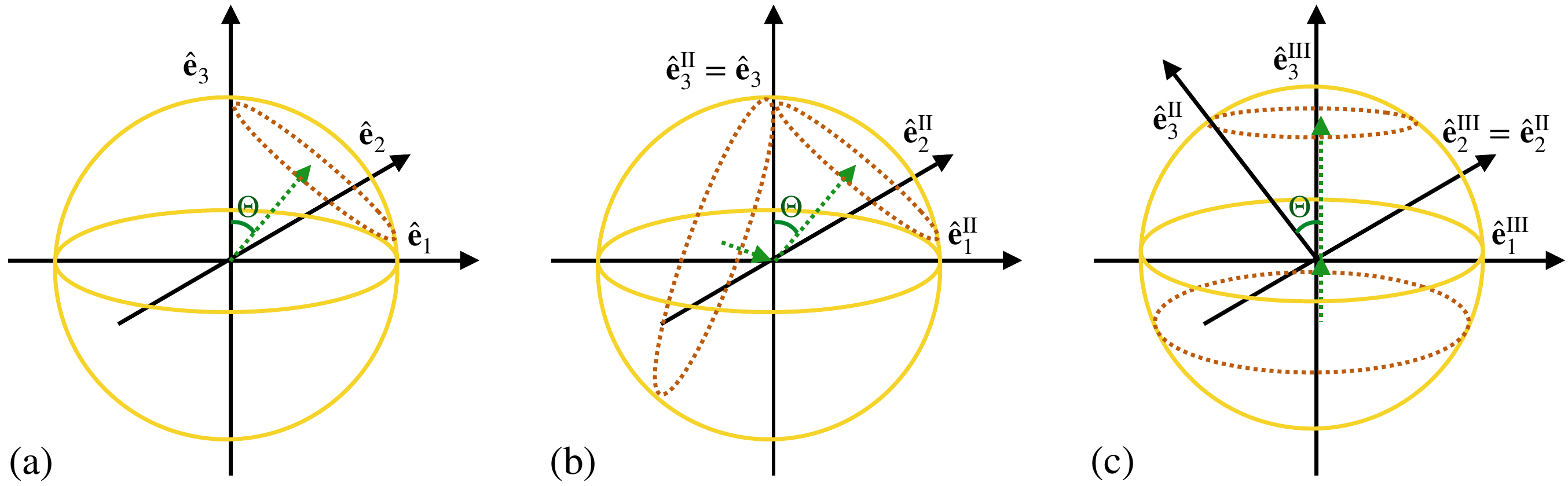}
	\caption{\label{fig:periodic_bipolar}Schematic diagrams for periodic bipolar motion on the Bloch sphere in three different frames. Brown dotted circles indicate the distributions of polarization vectors at several different times. Green dotted arrows indicate the central axis of these distributions. See text for details.}
\end{figure*}

The symmetry of the distribution of $\mathsf Q$ for the unstable eigenmode plays a crucial role in driving the periodic bipolar flavor evolution.
In the linear regime, they grow exponentially without being constrained by the curvature of the Bloch sphere because the sphere is locally flat in this limit.
However, when the system transitions into the non-linear regime, the fixed curvature of the Bloch sphere may force the distribution of $\mathsf P_\perp$ to deform from the initial shape of $\mathsf Q$, unless $\mathsf Q$ initially lie on a circle.
The circular symmetry in the distribution of $\mathsf Q$ can arise if their corresponding Hamiltonian vectors can be expressed in the form of $\mathbf{X}+u \mathbf{Y}$\footnote{We use the notation $\mathbf{X}$ and $\mathbf{Y}$ instead of the specific $\mathbf B$ or $\mu\mathbf J^\rho$ to emphasize the generality of this Hamiltonian form here.}.
The variable $u$ can be any arbitrary function of $\omega$, $v^x$, $v^y$, and $v^z$, as long as the Hamiltonian vectors form a one-dimensional linear distribution in the flavor space.
For any such Hamiltonian vectors, $\mathsf Q$ in Eq.~\eqref{eq:cp_ansatz} corresponds to
\begin{equation}
	\mathsf P_\perp(u) = \frac{\mathsf X_\perp+u\mathsf Y_\perp}{X_3 + u Y_3-\Omega},
    \label{eq:bipolar_unstable_mode}
\end{equation}
where $\mathsf P_\perp$, $\mathsf X_\perp$, and $\mathsf Y_\perp$ contain the same time-dependent factor $e^{-i\Omega t}$.
For $Y_3\neq 0$, the above configuration forms a \textit{circular} distribution centered at $\mathsf O = i \mathsf X_\perp /(2\Omega_i)  + [\Omega_i+i(\Omega_r-X_3)] \mathsf Y_\perp/(2\Omega_i Y_3) $ on the complex plane.
Details of the proof are given in Appendix~\ref{sec:circular_unstable}.

The Hamiltonian vector in Eq.~\eqref{eq:pvw_vdrho_P} may not be written in the form of $\mathbf{X}+u\mathbf{Y}$ in general. 
We discuss two special cases where this form applies.
For the single-angle slow mode with $\mathbf{H}=\omega\mathbf{B}+\lambda^t\hat{\mathbf e}_3+\mu\mathbf{J}^t$, we have $u=\omega$, $\mathbf{X}=\lambda^t\hat{\mathbf e}_3+\mu \mathbf{J}^t$, and $\mathbf{Y}=\mathbf{B}$.
For the single-energy AS fast mode with $\mathbf{H}=\lambda^t\hat{\mathbf e}_3+\mu(\mathbf{J}^t-v^z\mathbf{J}^z)$, we have $u=v^z$, $\mathbf{X}=\lambda^t\hat{\mathbf e}_3+\mu \mathbf{J}^t$, and $\mathbf{Y}=-\mu \mathbf{J}^z$. 
A special and important property of these two modes is that the distribution of $\mathsf P_\perp$ in Eq.~\eqref{eq:bipolar_unstable_mode} forms a circle that \textit{passes through the origin} of the complex plane, which requires $\mathrm{Im}(\mathsf X_\perp \mathsf Y_\perp^*) \approx 0$ (see Appendix~\ref{sec:circular_unstable}).
This condition is satisfied because $\mathsf Y_\perp=\mathsf{B}_\perp \approx 0$ for the single-angle slow mode and $\mathsf X_\perp=\mu\mathsf{J}^t_\perp \approx 0$\footnote{From Eq.~\eqref{eq:pvw_vdrho_P} it can be shown that $\partial_t \mathbf J^t = \lambda^t \hat{\mathbf e}_3 \times \mathbf J^t$. For $\mathbf J^t$ parallel to $\hat{\mathbf e}_3$ in the initial state, it remains so during subsequent evolution.} for the single-energy AS fast mode.
Note that for the latter mode, the circle for $\mathsf P_\perp$ passes through the origin at $u=0$.

How the periodic bipolar solution arises from the above properties of the unstable eigenmode $\mathsf{Q}$ can be understood by connecting the motion in the linear and non-linear regimes as follows.
In the linear regime, the configuration of $\mathsf P_\perp$ on the complex plane corresponds to that of vectors $\mathbf P_\perp$ on the horizontal $\hat{\mathbf e}_1$--$\hat{\mathbf e}_2$ plane.
Because $|\mathsf P_\perp| \ll P_3 \approx 1$, this circular configuration of $\mathbf P$ almost lies horizontally on the Bloch sphere, which is locally flat.
The origin of the corresponding complex plane is at the point $(0,0,1)$ in the full flavor space spanned by $\{\hat{\mathbf e}_1,\hat{\mathbf e}_2,\hat{\mathbf e}_3\}$.
The time-dependent factor $e^{-i\Omega t}=e^{-i\Omega_r t}e^{\Omega_i t}$ in $\mathsf P_\perp$ implies that the polarization vectors are rotating around $\hat{\mathbf e}_3$ with the angular speed $\Omega_r$ while the radius of their circular configuration is expanding exponentially with the rate constant $\Omega_i$.
Because the corresponding circle always passes through the point $(0,0,1)$, the overall motion can be captured by the angle $\Theta$ between the vertical direction ($\hat{\mathbf e}_3$) and the axis passing through the center of the circle (hereafter the central axis).
For definiteness, the axis points upwards in the linear regime and its direction during subsequent evolution is such that $\Theta$ changes continuously.

Because the tips of $\mathbf P$ are confined to the surface of the Bloch sphere, their configuration can no longer lie in the horizontal plane when their horizontal components $\mathsf P_\perp$ grow sufficiently large (i.e., when the curvature of the Bloch sphere starts to matter).
We propose the following solution for the motion of $\mathbf P$ in the nonlinear regime.
The configuration of $\mathbf P$ still lies in a plane, which intersects the Bloch sphere to form a circle.
The circle also passes through the point $(0,0,1)$ and the angle $\Theta$ between its central axis and $\hat{\mathbf e}_3$ evolves continuously to large values [see Fig.~\ref{fig:periodic_bipolar}(a)].

The general evolution of $\mathbf P$ can be elucidated with the help of two frame transformations.
Frame II rotates with the angular velocity $\Omega_r\hat{\mathbf e}_3$ with respect to the original frame, and the central axis of the circle for $\mathbf{P}^\text{II}$ lies in the $\hat{\mathbf e}_1^\text{II}$--$\hat{\mathbf e}_3^\text{II}$ plane ($\hat{\mathbf e}_3^\text{II}=\hat{\mathbf e}_3$).
The EOM in Frame II is
\begin{equation}
\partial_t\mathbf{P}^\text{II}=\mathbf{H}^\text{II}\times\mathbf{P}^\text{II},
\end{equation}
where
\begin{equation}
	\mathbf{H}^\text{II} = \mathbf{X}^\text{II} + u \mathbf{Y}^\text{II} - \Omega_r \hat{\mathbf{e}}_3^\text{II}.
\end{equation}
Frame III rotates with the angular velocity $(\partial_t\Theta)\hat{\mathbf e}_2^\text{II}$ with respect to Frame II, and the central axis of the circle for $\mathbf{P}^\text{III}$ points in the direction of $\hat{\mathbf e}_3^\text{III}$.
With $\hat{\mathbf{e}}_3^\text{II}=\cos\Theta\hat{\mathbf{e}}_3^\text{III}-\sin\Theta\hat{\mathbf{e}}_1^\text{III}$ and $\hat{\mathbf{e}}_2^\text{III}=\hat{\mathbf{e}}_2^\text{II}$ [see Figs.~\ref{fig:periodic_bipolar}(b) and \ref{fig:periodic_bipolar}(c)], the EOM in Frame III is
\begin{equation}
\partial_t\mathbf{P}^\text{III}=\mathbf{H}^\text{III}\times\mathbf{P}^\text{III},
\end{equation}
where 
\begin{align}
	\mathbf{H}^\text{III} = & \mathbf{X}^\text{III}
	+ u \mathbf{Y}^\text{III} 
	- \Omega_r \cos \Theta\, \hat{\mathbf{e}}_3^\text{III} \nonumber\\
	& + \Omega_r \sin \Theta\, \hat{\mathbf{e}}_1^\text{III}
	- (\partial_t \Theta)\, \hat{\mathbf{e}}_2^\text{III}.
    \label{eq:bipolar_pv_H_rr}
\end{align}
In terms of $\mathsf P_\perp^\text{III}$, the EOM is
\begin{align}
	i\partial_t \mathsf P_\perp^\text{III} = &
	[X_3^\text{III} + u Y_3^\text{III} - \Omega_r \cos\Theta] \mathsf P_\perp^\text{III} \nonumber\\
	 & -P_3^\text{III} [\mathsf X^\text{III}_\perp + u \mathsf Y^\text{III}_\perp + \Omega_r\sin\Theta +i\partial_t\Theta].
	 \label{eq:bipolar_pv_rr}
\end{align}

The evolution of $\mathbf{P}^\text{II}$ and $\mathbf{P}^\text{III}$ is visualized in Fig.~\ref{fig:periodic_bipolar}(b) and \ref{fig:periodic_bipolar}(c), respectively.
In Frame II, the circle for $\mathbf{P}^\text{II}$ hangs from the point $(0,0,1)$ and swings clockwise ($\Theta$ increasing from 0 to $\pi$) for a complete cycle.
The radius of the circle changes as it swings so that the circle passes $(0,0,-1)$ when $\Theta=\pi/2$ and returns back to $(0,0,1)$ when $\Theta=\pi$. 
In Frame III, the circle for $\mathbf{P}^\text{III}$ is always horizontal. It is initially near the north pole of the Bloch sphere. 
As it drops downwards, its radius first expands in the upper hemisphere and then shrinks after crossing the equator.

It is clear from the above discussion (see also Fig.~\ref{fig:periodic_bipolar}) that $|\mathsf P_\perp^\text{III}|=\sin\Theta$ and $P_3^\text{III}=\cos\Theta$. 
For both the single-angle slow and single-energy AS fast modes, $X_3^\text{III}=\tilde X_3\cos\Theta$ and $Y_3^\text{III}=\tilde Y_3\cos\Theta$, where the tilde symbol denotes quantities evaluated for $P_3=1$.
In order to make the solution complete, we set $\partial_t \Theta = \Omega_i \sin\Theta$ and rewrite Eq.~\eqref{eq:bipolar_pv_rr} as
\begin{align}
	i\partial_t \mathsf P_\perp^\text{III} = &
	\cos\Theta\{[\tilde X_3 + u\tilde Y_3 - \Omega_r]\mathsf P_\perp^\text{III} \nonumber\\
	 & -[\mathsf X^\text{III}_\perp + u \mathsf Y^\text{III}_\perp + (\Omega_r+i\Omega_i)|\mathsf P_\perp^\text{III}|]\}.
	 \label{eq:bipolar_pv_rr2}
\end{align}
The above choice of $\partial_t \Theta$ can be justified by comparing Eq.~\eqref{eq:bipolar_pv_rr2} to the EOM in the linear regime, which corresponds to $|\mathsf P_\perp^\text{III}|\approx\Theta\ll 1$.
 
Recall that $i\partial_t\mathsf P_\perp=(\Omega_r+i\Omega_i)\mathsf P_\perp$ in the linear regime.
With $\mathsf P_\perp=\mathsf P_\perp^\text{II}e^{-i\Omega_rt}$, we obtain $\partial_t\mathsf P_\perp^\text{II}=\Omega_i\mathsf P_\perp^\text{II}$ and therefore, $\partial_t\mathsf O^\text{II}=\Omega_i\mathsf O^\text{II}$ for the center of the circle for $\mathsf P_\perp^\text{II}$.
The EOM for $\mathsf P_\perp^\text{II}$ is
\begin{equation}
	i\partial_t \mathsf P^\text{II}_\perp =
	H_3^\text{II} \mathsf P^\text{II}_\perp - \mathsf H^\text{II}_\perp, 
   \label{eq:bipolar_cp_dtPperp_r}
\end{equation}
where $H^\text{II}_3=\tilde X_3+u \tilde Y_3-\Omega_r$ and $\mathsf H^\text{II}_\perp=\mathsf X^\text{II}_\perp + u\mathsf Y^\text{II}_\perp$.
The transformation from Frame II to III in the linear regime ($\mathsf O^\text{II}\ll 1$) is equivalent to $\mathsf P_\perp^\text{III}=\mathsf P_\perp^\text{II}-\mathsf O^\text{II}$, which along with Eq.~\eqref{eq:bipolar_cp_dtPperp_r} gives\footnote{For both the single-angle slow and single-energy AS fast modes, $\mathsf X^\text{III}_\perp$ and $\mathsf Y^\text{III}_\perp$ contain some terms proportional to $\mathsf O^\text{II}$ due to the transformation $\mathsf P_\perp^\text{III}=\mathsf P_\perp^\text{II}-\mathsf O^\text{II}$.}
\begin{align}
	i\partial_t \mathsf P^\text{III}_\perp = i\Omega_i\mathsf P_\perp^\text{III}&=
	[\tilde X_3+u \tilde Y_3-\Omega_r]\mathsf P^\text{III}_\perp \nonumber\\
    &-[\mathsf X^\text{III}_\perp + u\mathsf Y^\text{III}_\perp + (\Omega_r + i\Omega_i)\mathsf O^\text{II}].
	\label{eq:bipolar_cp_rr}
\end{align}
Because the circle for $\mathsf P_\perp^\text{II}$ passes through the point $(0,0,1)$ in Frame II, $|\mathsf P_\perp^\text{III}|=\mathsf O^\text{II}$. 
Comparing Eqs.~\eqref{eq:bipolar_pv_rr2} and \eqref{eq:bipolar_cp_rr}, we not only see that they agree in the limit $\Theta\ll1$, but also obtain for the nonlinear regime
\begin{equation}
	\partial_t \mathsf P_\perp^\text{III} = \Omega_i \mathsf P_\perp^\text{III} \cos \Theta,
\end{equation}
which agrees with our solution of expanding or contracting circles for $\mathsf P_\perp^\text{III}$.
Note that the growth rate of these circles is related to the angular speed $\partial_t \Theta$ for rotation of Frame III relative to Frame II.

From $\partial_t \Theta = \Omega_i \sin\Theta$, we obtain
\begin{equation}
    \Theta(t) = 2 \arctan [e^{\Omega_i(t-t_0)}],
    \label{eq:periodic_Theta}
\end{equation}
where $t_0$ is a reference time.
Therefore, the evolution of the unstable eigenmode for periodic bipolar motion in the nonlinear regime is analytically solvable, and the solution for polarization vectors with different $u$ can be readily obtained from Eqs.~\eqref{eq:periodic_Theta} and \eqref{eq:Appendix_B3}--\eqref{eq:Appendix_B5}.

If the circular distribution of the polarization vectors does not go through the point $(0,0,1)$ in Frame II, but somehow corresponds to $|\mathsf P^\text{III}_\perp|=\sin\Theta'$ and $P_3^\text{III}=\cos\Theta'$ for all $u$ in Frame III with $\Theta'\neq \Theta$, then terms associated with polarization currents in the first term on the right-handed side of Eq.~\eqref{eq:bipolar_pv_rr} contain $\cos \Theta'$ but other terms such as $\mathbf B^{\rm III}$ and $\Omega_r \cos\Theta$ contain $\cos \Theta$.
Consequently, there is no simple multiplicative relation between the EOM in the nonlinear regime and that in the linear one, which would most likely cause deviations of the flavor evolution from the perfect periodic bipolar solution.

Note that our picture of circular symmetry is \blue{consistent with} the pendulum formalism constructed by using three arbitrary $v^z$ \cite{padilla2022neutrino}.
Because at least three points are needed to form a unique circle, the circular distribution of all polarization vectors can be specified using three different values of $v^z$.
The evolution for other $v^z$ can then be determined based on Eq.~\eqref{eq:Appendix_B3}.

\section{From periodic to aperiodic}
\label{sec:aperioic}
\begingroup
\begin{table*}[t]
\begin{ruledtabular}
    \caption{\label{tab:models} Spectral and angular distributions $F_{\omega,\vec v}$ as well as discretization schemes on $\omega$, $v^z$, and the azimuthal angle $\phi$ in $v^x$--$v^y$ plane for all eight models. 
    Letter ``Y'' (``N'') in the columns of $H_{\rm vac}$ or $H_{\rm mat}$ indicates that these terms are (not) included.
    If $H_{\rm mat}$ is included we take $\lambda^\rho=(\mu,\,0,\,0,\,0.5\mu)$. 
    For cases where $H_{\rm vac}$ is included, the neutrino mass ordering is taken to be inverted.
    For the discretization schemes, unless specified by fixed values, $\omega$, $v^z$, and $\phi$ are discretized uniformly in a range by the given number of bins. The function $g$ is defined in Eq.~\eqref{eq:angular_distribution_g}.}
    \centering
    \begin{tabular}{cc cccc}
        model & $N_{\rm beam}$ & $H_{\rm vac}$ & $H_{\rm mat}$ & discretization schemes & $F_{\omega,\vec v}$  \\\hline
        Two-beam slow & 2 & Y & N &  \makecell{$v^x=v^y=v^z=0$ effectively; \\ $\omega$ is either 0.1$\mu$ or $-0.1\mu$} & $F_{\omega} = {\rm sgn}(\omega)+0.5$ \\\hline
        Single-angle slow & 10000 & Y & N & \makecell{$v^x=v^y=v^z=0$ effectively; \\ 10000 bins for $-0.2\mu<\omega<0.2\mu$} & $F_{\omega} = {\rm sgn}(\omega)+0.5$ \\\hline
        Single-energy AS fast & 10000 & N & N & \makecell{$v^x=v^y=0$ effectively; \\ 10000 bins for $-1<v^z<1$} & $F_{v^z} = g(v^z, 0.9) $  \\\hline
        \makecell{Four-beam \\ coplanar fast \cite{dasgupta2018fast,sawyer2021neutrino}} & 4 & N & N & \makecell{$v^z=0$; $\phi$ takes $\pi/6,$ $5\pi/6,$ \\ $7\pi/6,$ and $11\pi/6$ respectively} & $F_{v^x,v^y}={\rm sgn}(v^y)$ \\\hline
        \makecell{Eight-beam \\ coplanar fast} & 8 & N & N & \makecell{$v^z=0$; \\ 8 bins for $0<\phi<2\pi$} & $F_{v^x,v^y}={\rm sgn}(v^y)$ \\\hline
        \makecell{AS fast with non-zero \\ matter bulk velocity \cite{padilla2021fast}} & 10000 & N & Y & \makecell{$v^x=v^y=0$ effectively; \\ 10000 bins for $-1<v^z<1$} & $ F_{v^z} = g \left( v^z, 0.9 \right) $ \\\hline
        MZA slow & 40000 & Y & N & \makecell{$v^x=v^y=0$ effectively; \\ 200 bins for $-0.2\mu<\omega<0.2\mu$; \\ 200 bins for $-1<v^z<1$} & \makecell{$F_{\omega, v^z} = [{\rm sgn}(\omega)+0.5]$ \\ $\times (1+0.5v^z)$} \\\hline
        SB fast \cite{xiong2021stationary} & 38400 & N & N & \makecell{300 bins for $-1< v^z< 1$; \\ 128 bins for $0<\phi<2\pi$} & $F_{v^x,v^y,v^z} = g(v^z, 1.1)$ \\
    \end{tabular}
\end{ruledtabular}
\end{table*}
\endgroup

We use eight representative models covering most of the homogeneous models with a single spectral crossing in recent years to illustrate the association between the symmetry of the unstable eigenmode and the features of flavor evolution.
They are implemented numerically so that all integrals are replaced by the sum over discretized beams with respect to $\omega$ and $\vec v$.
Equation~\eqref{eq:pvw_vdrho_P} becomes
\begin{equation}
	\partial_t \mathbf{P}_{\omega,\vec{v}} = \mathbf{H}_{\omega,\vec{v}} \times \mathbf{P}_{\omega,\vec{v}},
    \label{eq:EOM_discretized}
\end{equation}
where
$ \mathbf{H}_{\omega,\vec{v}} = \omega \mathbf{B} + (v^\rho)_{\vec{v}} ( \lambda_\rho \hat{\mathbf{e}}_3 + \mu \mathbf{J}_\rho) $,
$ \mathbf{J}^\rho = N_\mathrm{beam}^{-1} \sum_{\omega, \vec v}^{N_\mathrm{beam}} (v^\rho)_{\vec{v}} F_{\omega,\vec{v}} \mathbf{P}_{\omega,\vec{v}} $, and $N_\mathrm{beam}$ is the total number of beams.
We take $F_{\omega,\vec{v}}=N_\mathrm{beam} F(\omega,\vec{v}) \Delta \omega \Delta^2 \vec v$ with the grid widths $\Delta \omega$ and $\Delta \vec v$ so that the value of $F_{\omega,\vec{v}}$ is independent of how many beams are discretized.
The corresponding linearized EOM is
$ \Omega \mathsf Q_{\omega,\vec{v}} =
	(H_3)_{\omega,\vec{v}} \mathsf Q_{\omega,\vec{v}}
	- \mathsf T_{\omega,\vec{v}} $,
with $\mathsf T_{\omega,\vec{v}} = \mu N_\mathrm{beam}^{-1} (v_\rho)_{\vec v} \sum_{\omega', \vec v'}^{N_\mathrm{beam}} (v^\rho)_{\vec{v}'} F_{\omega',\vec{v}'} \mathsf Q_{\omega',\vec{v}'}$ and the unstable eigenmode $\mathsf Q_{\omega,\vec{v}}=\mathsf T_{\omega,\vec{v}}/[(H_3)_{\omega,\vec{v}}-\Omega]$.

Parameters and discretization schemes on $\omega$, $v^z$, and the azimuthal angle $\phi$ in $v^x$--$v^y$ plane for all eight models are listed in Table~\ref{tab:models} where 
\begin{equation}
g(v^z, \alpha)
= \frac{20}{\sqrt{\pi}} 
\left[ \sigma_\nu^{-1} e^{-(\frac{1-v^z}{\sigma_\nu})^2} 
-\alpha \sigma_{\bar\nu}^{-1} e^{-(\frac{1-v^z}{\sigma_{\bar\nu}})^2}  \right],
\label{eq:angular_distribution_g}
\end{equation}
$\sigma_\nu=0.6\sqrt{2}$, and $\sigma_{\bar\nu}=0.5\sqrt{2}$.

\subsection{Breaking-down of periodic bipolar motion}
\label{sec:breaking_down}
\begin{figure*}[hbt!]
	\centering
		\subfloat[Two-beam slow]{\includegraphics[width=0.25\textwidth]{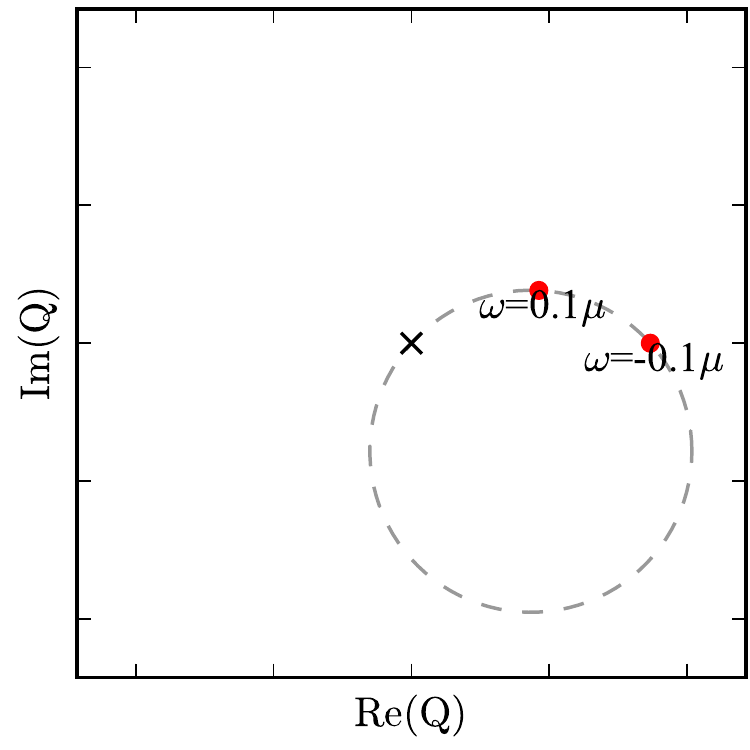}}
		\subfloat[Single-angle slow. Various colors are for different $\omega$.]{\includegraphics[width=0.25\textwidth]{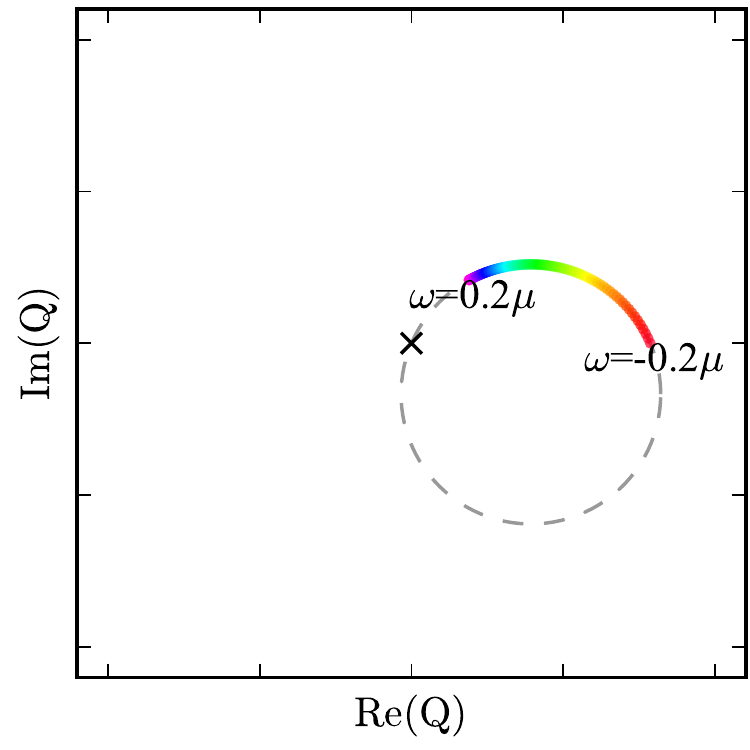}}
		\subfloat[AS fast. Various colors are for different $v^z$.]{\includegraphics[width=0.25\textwidth]{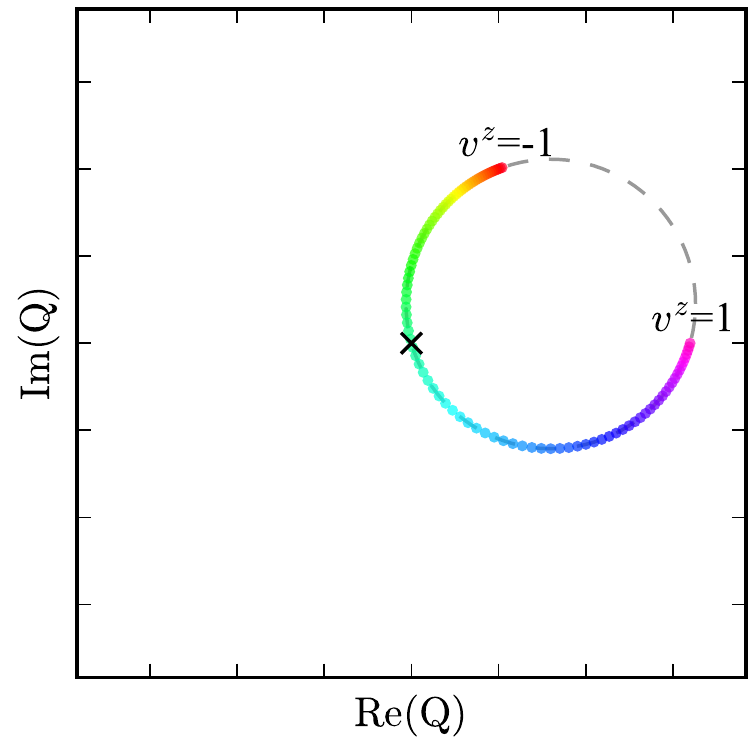}}
		\subfloat[Four-beam coplanar fast. Values of $\phi$ are marked.]{\includegraphics[width=0.25\textwidth]{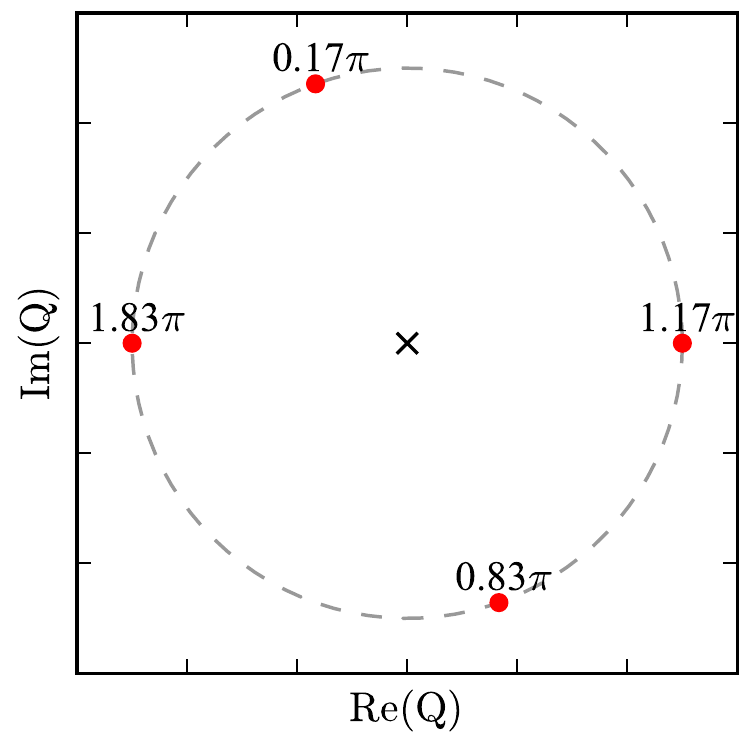}}\\
		\subfloat[Eight-beam coplanar fast. Values of $\phi$ are marked.]{\includegraphics[width=0.25\textwidth]{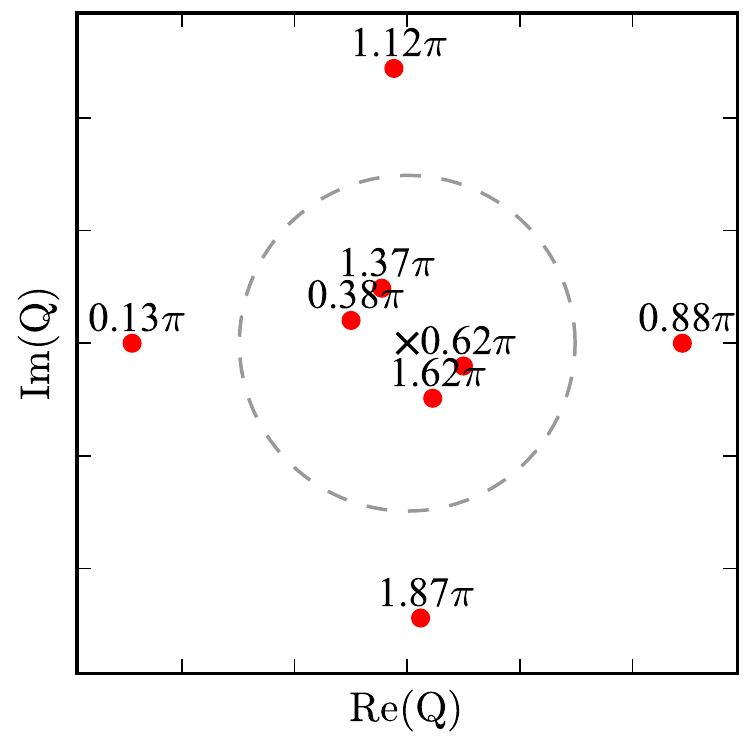}}
		\subfloat[AS fast with non-zero bulk velocity \blue{of} matter. Various colors are for different $v^z$.]{\includegraphics[width=0.25\textwidth]{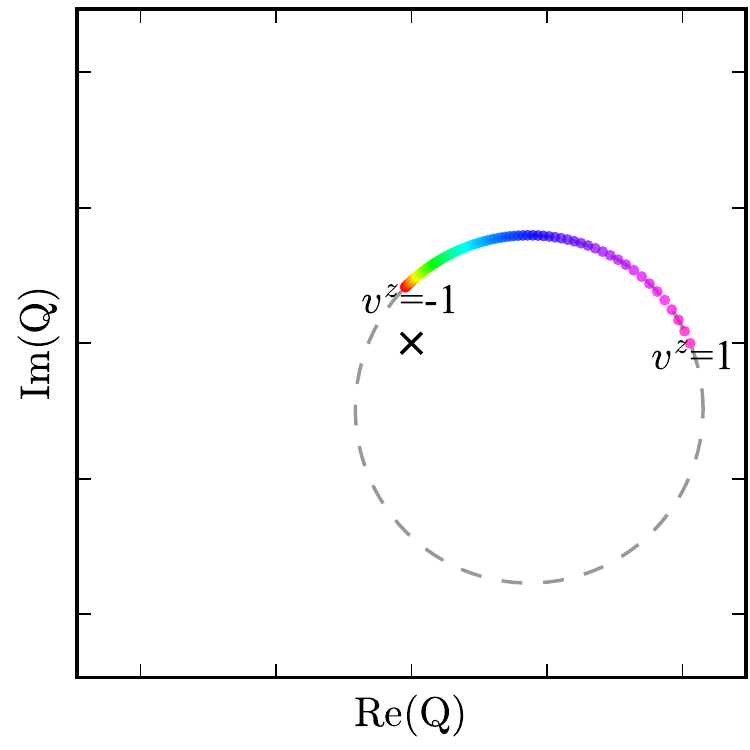}}
		\subfloat[MZA slow. Various colors and marker sizes are for different $\omega$ and $v^z$ respectively.]{\includegraphics[width=0.25\textwidth]{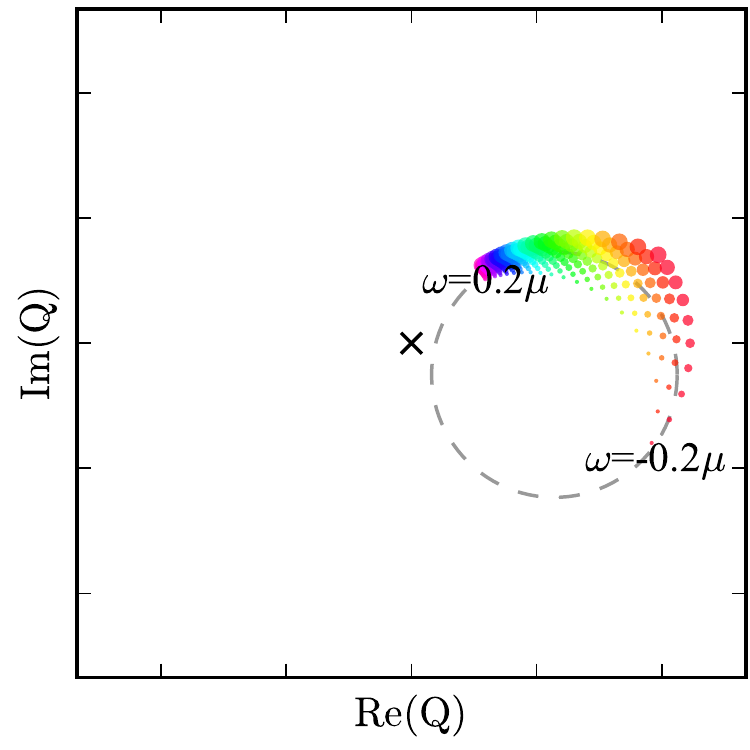}}
		\subfloat[SB fast. Various colors and marker sizes are for different $\phi$ and $v^z$ respectively.]{\includegraphics[width=0.25\textwidth]{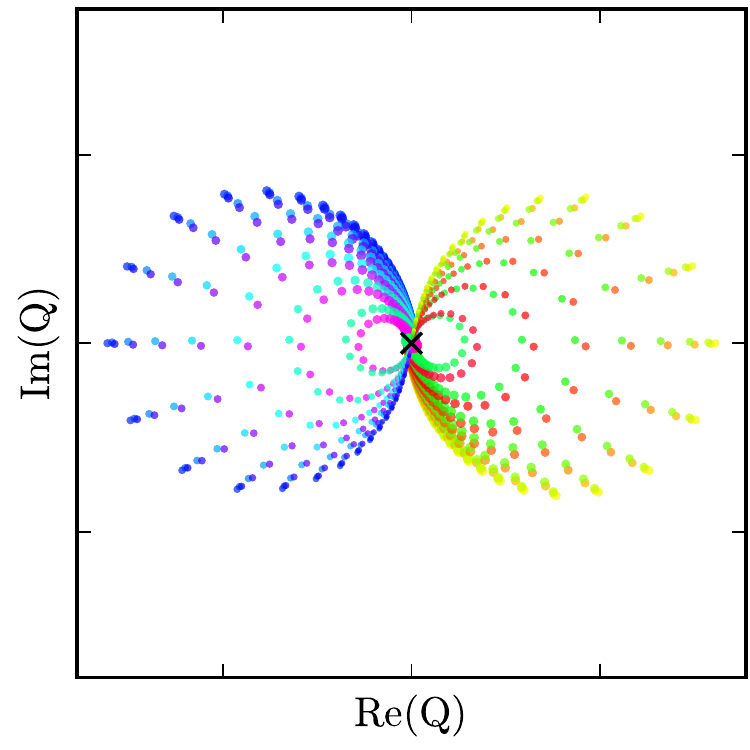}}
	\caption{\label{fig:unstable_mode} Distributions of unstable eigenmodes $\mathsf Q$ on complex planes for all eight models. 
    Gray dashed circles are shown in panels~(a)--(g) to illustrate how close they are to circular distributions.
    The cross at the center of each panel marks the origin of the complex plane.
    Because unstable eigenmodes can be rescaled arbitrarily, we do not show numerical values on the axes except for ticks marking equal intervals.}
\end{figure*}

\begin{figure*}[hbt!]
	\centering
		\subfloat[Two-beam slow]{\includegraphics[width=0.25\textwidth]{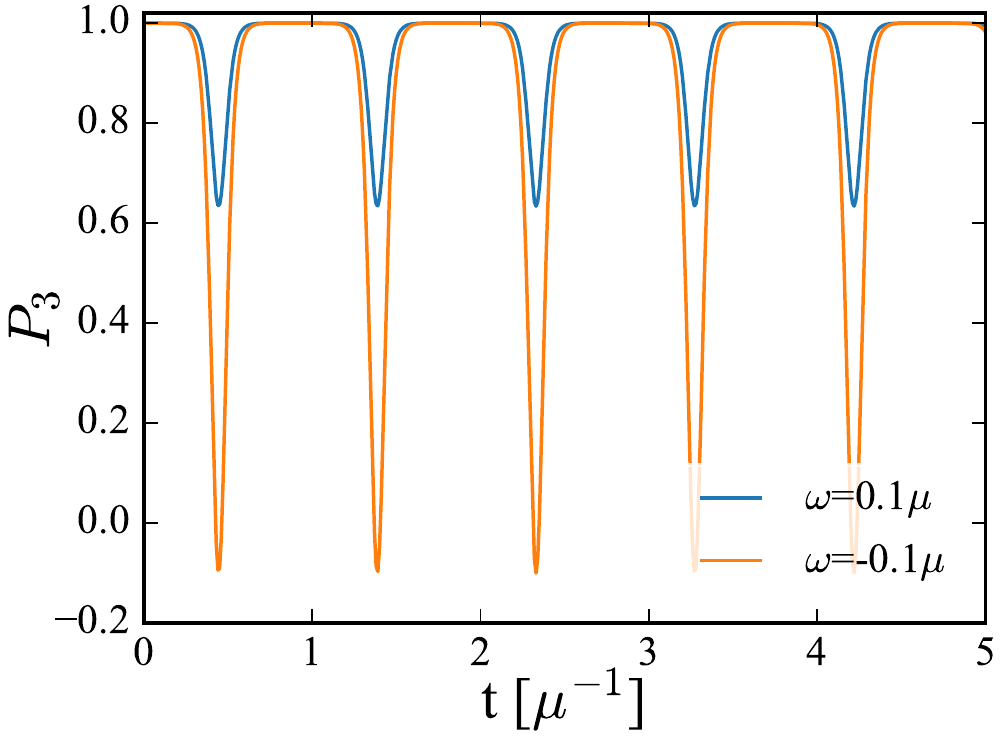}}
		\subfloat[single-angle slow]{\includegraphics[width=0.25\textwidth]{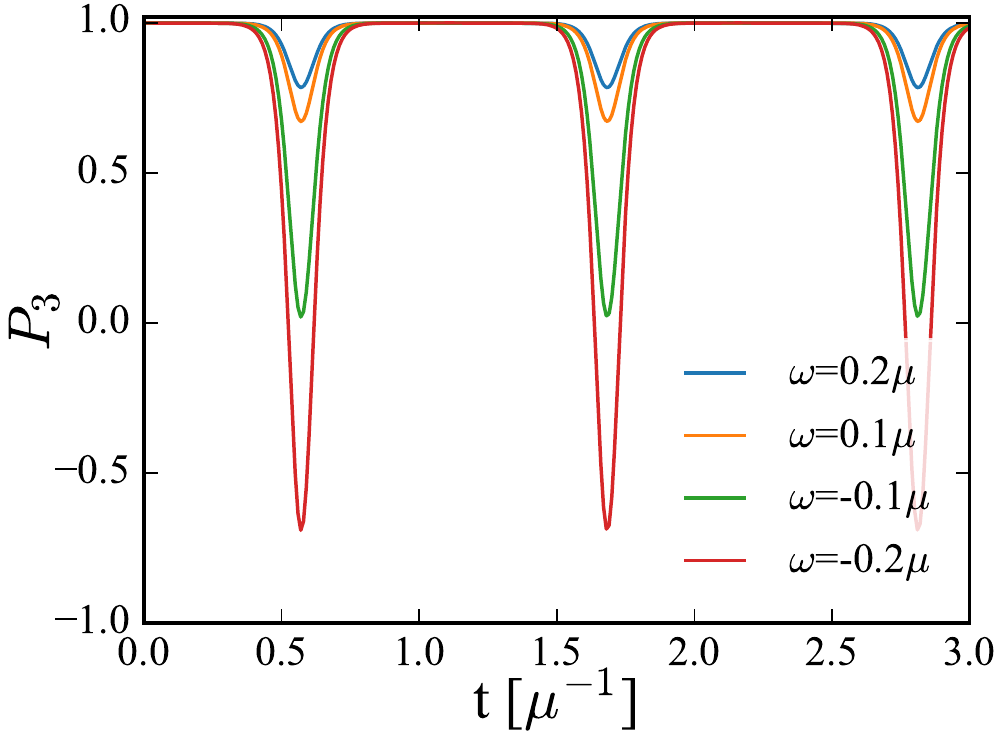}}
		\subfloat[AS fast]{\includegraphics[width=0.25\textwidth]{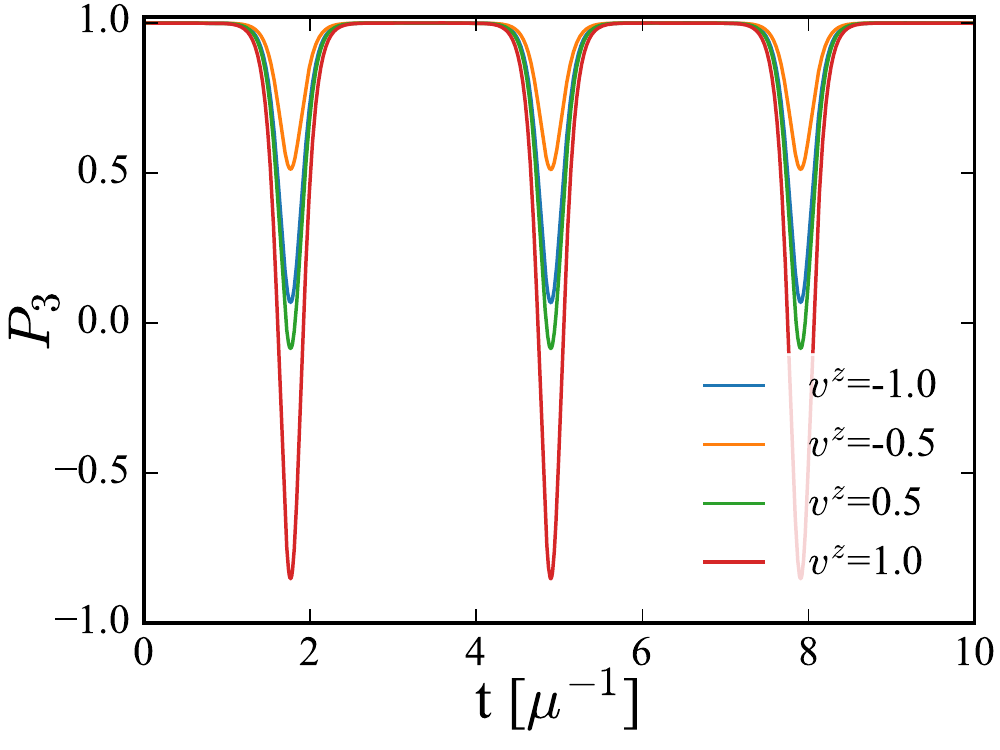}}
		\subfloat[four-beam coplanar fast]{\includegraphics[width=0.25\textwidth]{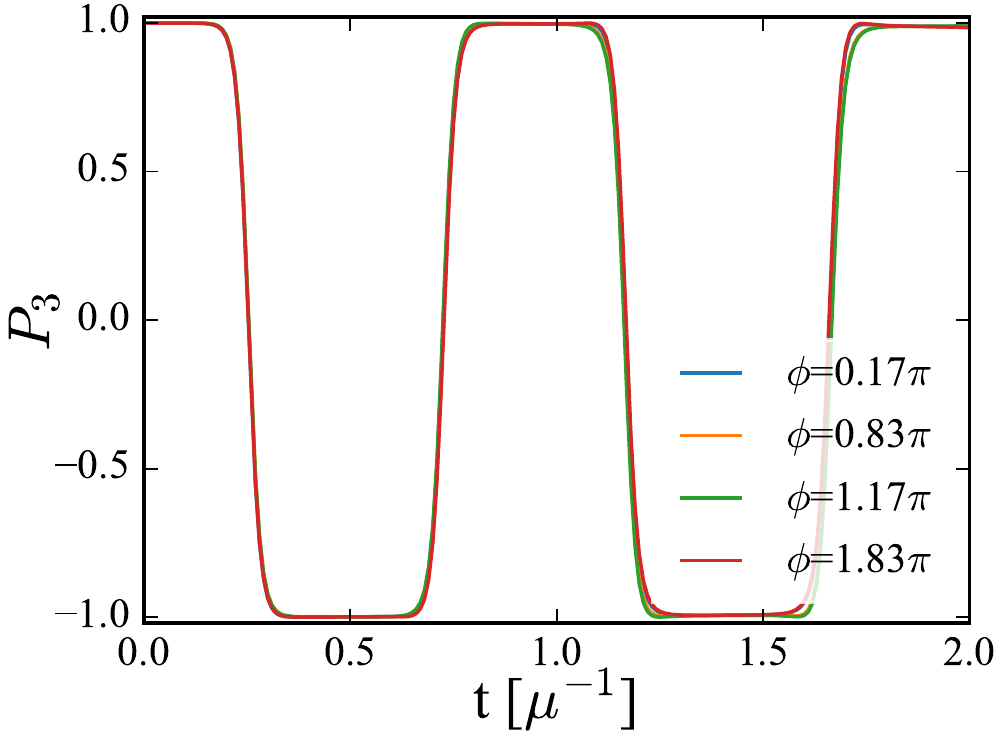}}\\
		\subfloat[eight-beam coplanar fast]{\includegraphics[width=0.25\textwidth]{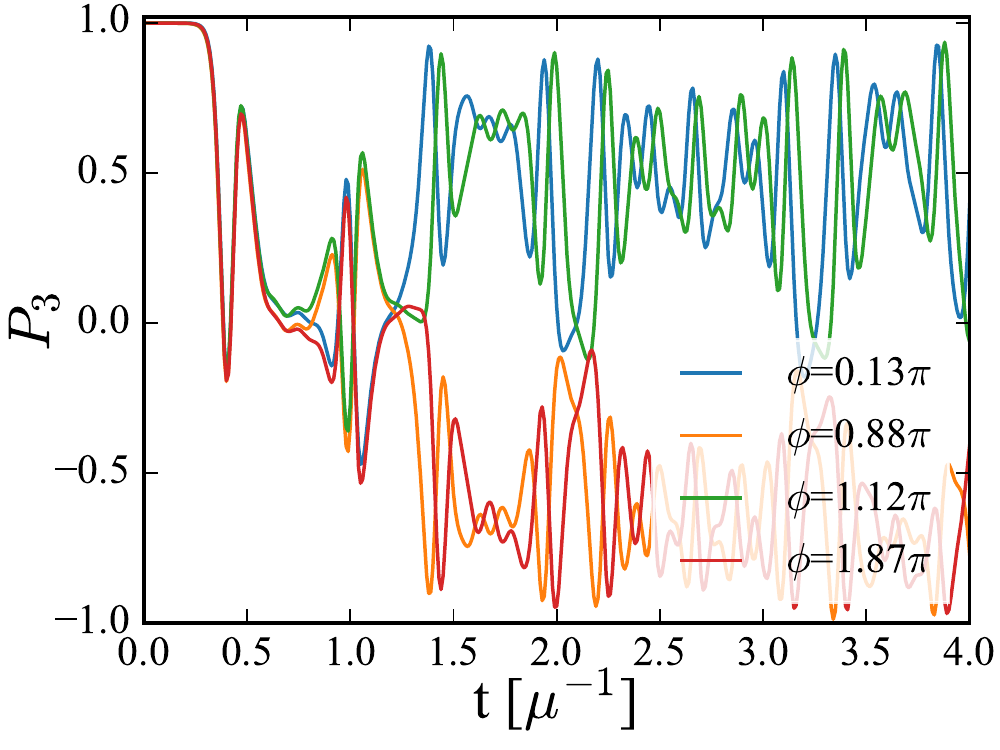}}
		\subfloat[AS fast with bulk velocity]{\includegraphics[width=0.25\textwidth]{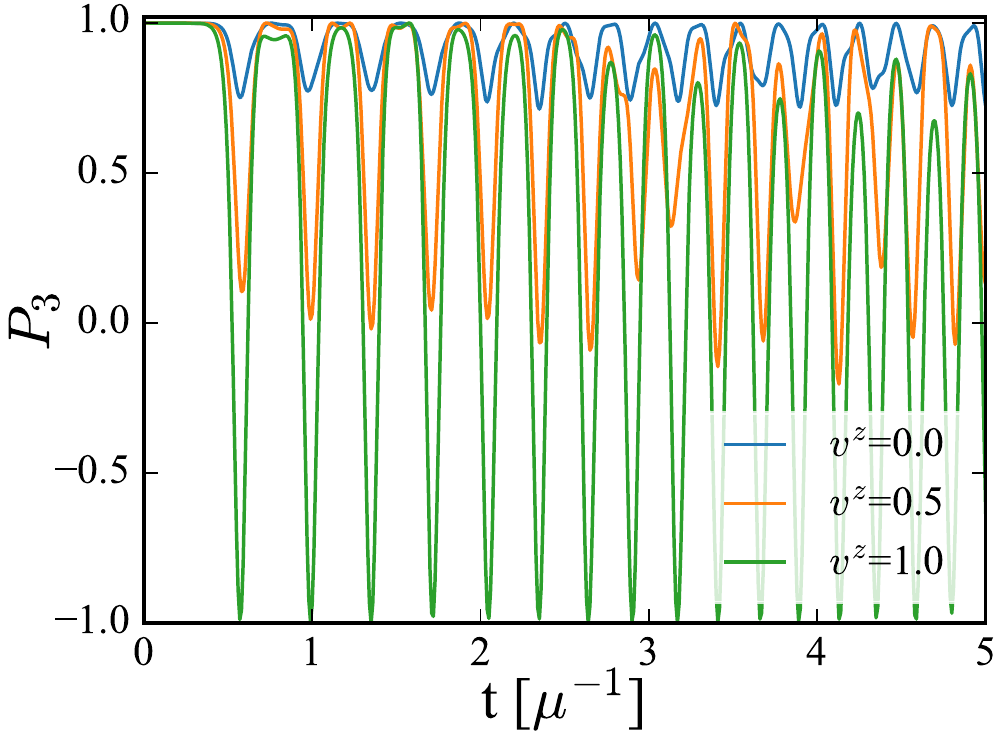}}
		\subfloat[MZA slow]{\includegraphics[width=0.25\textwidth]{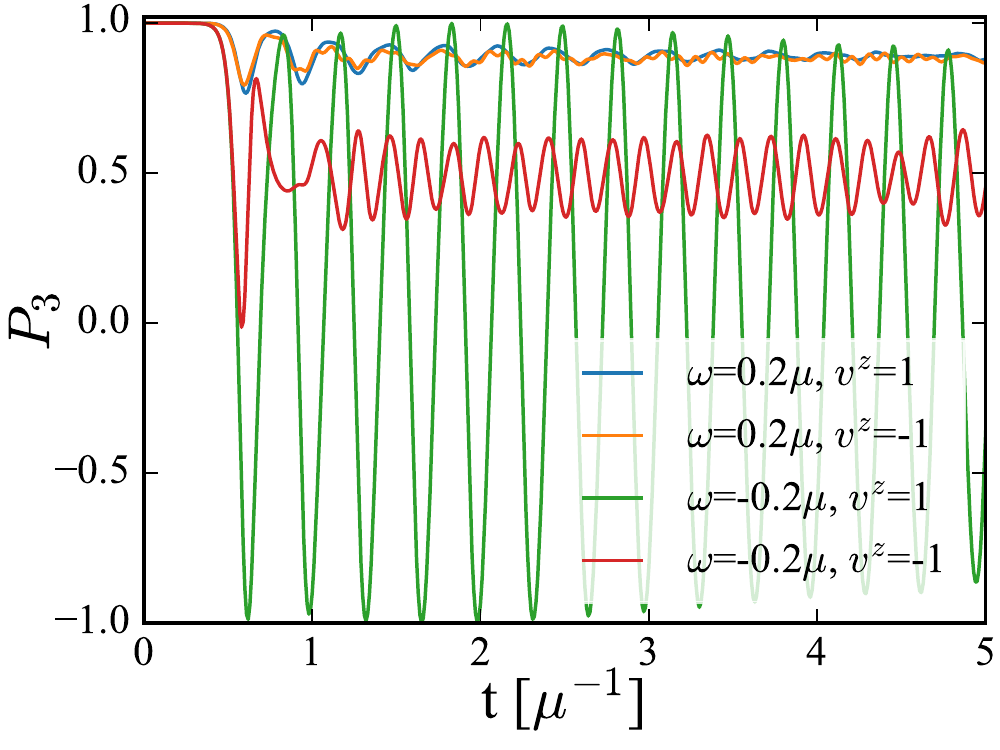}}
		\subfloat[SB fast]{\includegraphics[width=0.25\textwidth]{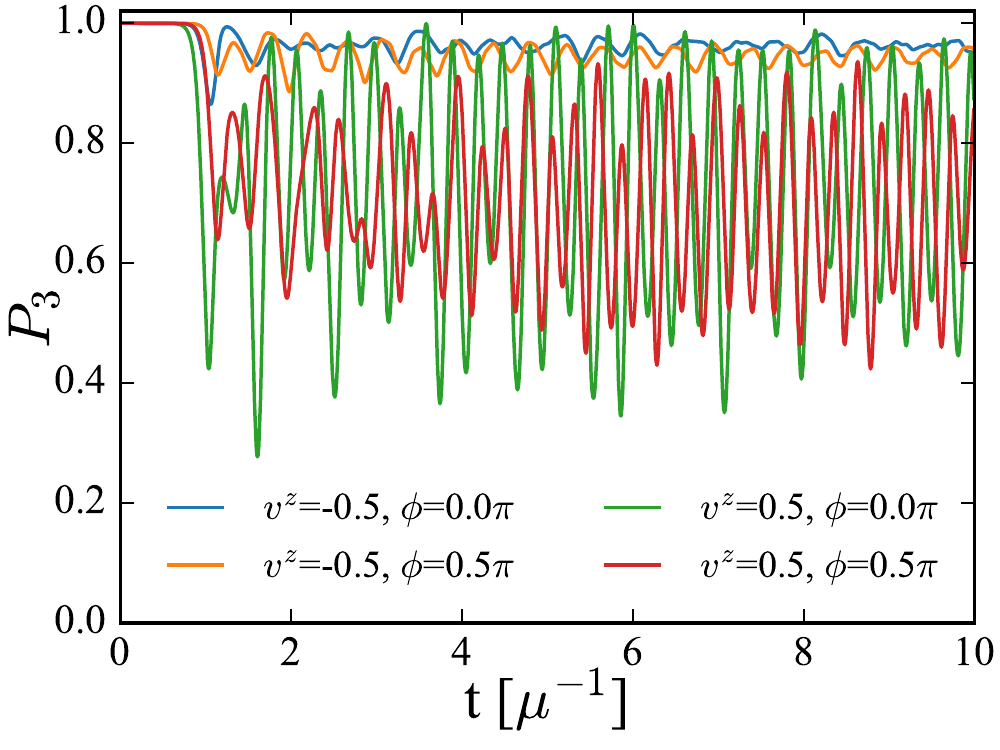}}
	\caption{\label{fig:evolution} Time evolution of $P_3$ in several representative beams for all eight models. Colors show different beams with descriptions in the legend of each panel.}
\end{figure*}

Figure~\ref{fig:unstable_mode} shows the distribution of $\mathsf{Q}$ of the unstable eigenmode on the complex plane for the above eight models.
For each model we also calculate the flavor evolution by solving Eq.~\eqref{eq:EOM_discretized} with an initial condition of $\mathbf P_{\omega,\vec v}$ perturbed from $\hat{\mathbf e}_3$ by a random deviation.
The random perturbation is implemented in the following way.
For each $\omega$ or $\vec v$, each of the three components in $\mathbf P_{\omega,\vec v}$ is added by $\delta_{\rm pert} \epsilon_{\rm pert}$ individually where $\delta_{\rm pert}=10^{-3}$ and $\epsilon_{\rm pert}$ is a random seed following a uniform distribution between $-1$ and 1.
After adding this random contribution, $\mathbf P_{\omega,\vec v}$ is normalized for each $\omega$ or $\vec v$ as the initial condition.
Figure~\ref{fig:evolution} shows the flavor evolution of several representative $\omega$ or $\vec v$.
The distribution of $\mathsf Q$ for the first three models all lie on a circle that passes through the origin.
As expected, they also show clean bipolar flavor evolution. 
In particular, the first model has only two beams so the distribution automatically forms a circle together with the origin. 
Because the initial condition is randomly perturbed rather than specified as the exact unstable eigenmode, the evolution does not exactly follow Eq.~\eqref{eq:periodic_Theta} that extends to infinite time.
Every time when the circular distribution contracts to a similar magnitude as in the initial perturbation, those residuals inherited from the initial random pattern prevent further contraction, which in turn, starts the next cycle of the bipolar motion.

The four-beam coplanar model in Refs.~\cite{dasgupta2018fast,sawyer2021neutrino} provides an interesting insight.
It does not follow the previous patterns where the unstable eigenmode is a circle going through the origin.
Instead, the symmetry among those four beams in their $v^x$ and $v^y$ distribution forces the unstable eigenmode to have the same $|\mathsf Q_{v^x,v^y}|$ given that $|\mathsf Q_{v^x,v^y}|=|\mathsf Q_{-v^x,v^y}|$ and $|\mathsf Q_{v^x,v^y}|=|\mathsf Q_{v^x,-v^y}|$.
In this case, they lie directly on a circle centered at the origin \blue{in the absence of vacuum and matter terms} so that when transitioning into non-linear regime the transformation between Frame II and III discussed in Sec.~\ref{sec:periodic} is not needed, or equivalently, $\Theta'=\Theta=0$.
As a result, the solution in Frame II simply expands and contracts in the upper and lower Bloch hemispheres, which leads to the symmetric and ``box-like'' shape for $P_3$ that looks different from the bipolar pattern exhibited in the first three models.

Of course, this four-beam coplanar model is very special and highly relies on the discrete symmetry in the $v^x$--$v^y$ distribution.
Slightly relaxing this symmetry, e.g., doubling the number of beams as in the eight-beam coplanar model leads to an unstable eigenmode whose $\mathsf Q$ are distributed on two concentric circles with different $|\mathsf Q_{v^x,v^y}|$.
Those two concentric circles grow with the same speed in the linear regime. 
However, once evolving to non-linear regime, the outer circle gets more curved than the inner one on Bloch sphere, which makes them unable to keep the same distribution of the unstable eigenmode as in the linear regime.
Hence, the evolution does not follow any bipolar shape at all, as shown in Fig.~\ref{fig:evolution}(e).
It is interesting to note that although the unstable eigenmodes in four-beam and eight-beam coplanar models both break the spatial reflection symmetry along $v_x$, it does not lead to different evolution of $P_3$ in the four-beam coplanar model, but does so in the eight-beam model.

For the next three models, we discuss other cases with ``continuous'' neutrino spectrum.
Beyond the simple slow and AS fast models above, there can be more ways to break the geometric symmetry required for the bipolar motion.
As discussed in Sec.~\ref{sec:periodic}, a key condition for the bipolar motion is that the growing speed of the unstable circle needs to match the rotating speed of its central axis. 
This condition can be easily broken down by shifting the circular unstable eigenmode away from the origin.
This shift can be achieved even for the AS fast mode that is commonly thought as the paradigm of bipolar pendulum. 
For example, in the presence of a flowing bulk matter, neutrinos moving in different directions experience different effective matter potentials.
From Eq.~\eqref{eq:pvw_vdrho_P}, we have $\partial_t \mathbf J^t = \lambda_\rho \hat{\mathbf e}_3 \times \mathbf J^\rho$ instead of $\partial_t \mathbf J^t = \lambda^t \hat{\mathbf e}_3 \times \mathbf J^t$.
It cannot be assumed that $|\mathsf J^t_\perp|$ is negligible as the system grows in the linear regime because it is not conserved now.
The horizontal component $\mathsf J^t_\perp$ enters into Eq.~\eqref{eq:bipolar_unstable_mode}, and $\mathrm{Im}(\mathsf J^t_\perp \mathsf J^{z*}_\perp)$ is not necessarily zero.
As a result, the circular unstable eigenmode does not go through the origin.
After a few cycles, it eventually deviates from bipolar motion through kinematic decoherence.

Another way of breaking the requirement for bipolar motion is to have more than one variable dependence on $\omega$ and $\vec v$ so that the distribution of $\mathsf Q$ of the unstable eigenmode no longer forms a simple one-dimensional arc.
The MZA slow mode is a good example as it explicitly contains the $\omega$ and $v^z$ dependence that cannot be removed.
The unstable eigenmode in Fig.~\ref{fig:unstable_mode}(g) occupies a two-dimensional area in the complex plane.
Thus, it is impossible to embedded this shape onto the curved surface of Bloch sphere when transitioning into the non-linear regime.
Unless other hidden geometric symmetry exists, which is perhaps possible but very rare, the distribution will be distorted on the Bloch sphere during the evolution.
For this case, the evolution of polarization vectors then undergoes the kinematic decoherence \cite{raffelt2007self}, or dubbed differently as the relaxation and cascades in multi-energy fast mode \cite{johns2020fast}.

The same way of breaking the symmetry of the eigenmode in principle should apply to the multi-energy fast unstable eigenmode since it involves both $\omega$ and $v_z$.
However, as $\mathbf H_{\nu\nu}$ dominates over the energy-dependent $\mathbf H_{\rm vac}$, its eigenmode distribution mainly depends on $v^z$ and the dependence on $\omega$ can be treated perturbatively.
Consequently, the flavor evolution still follows closely the bipolar motion in short time scale defined by the fast instability.
For the SB fast mode, the azimuthal symmetry in $v^x$-$v^y$ plane can be broken so that the distribution of unstable eigenmode in complex plane depends on both $v^z$ and the azimuthal angle $\phi$ as shown in Fig.~\ref{fig:unstable_mode}(h).
This distribution covers a much wider region than the MZA slow mode and obviously deviates the most from a single circular shape.
As a result, the distribution gets distorted in the non-linear regime.
The flavor evolution quickly cascades into quasi-stationary state and exhibits aperiodic behaviors at later times as shown in Fig.~\ref{fig:evolution}(h).

From the last four models discussed above, we find an overall trend that the more deviated from the circular distribution the unstable eigenmode is, the more aperiodic the behavior of the flavor evolution will be, particularly in eight-beam coplanar and SB fast models.
The unstable eigenmode in AS fast model with bulk velocity is least deviated among those four models so that nearly bipolar motions at the very first few oscillation periods were obtained.

\begin{figure*}[hbt!]
	\centering
		\subfloat[AS fast]{\includegraphics[width=0.5\textwidth]{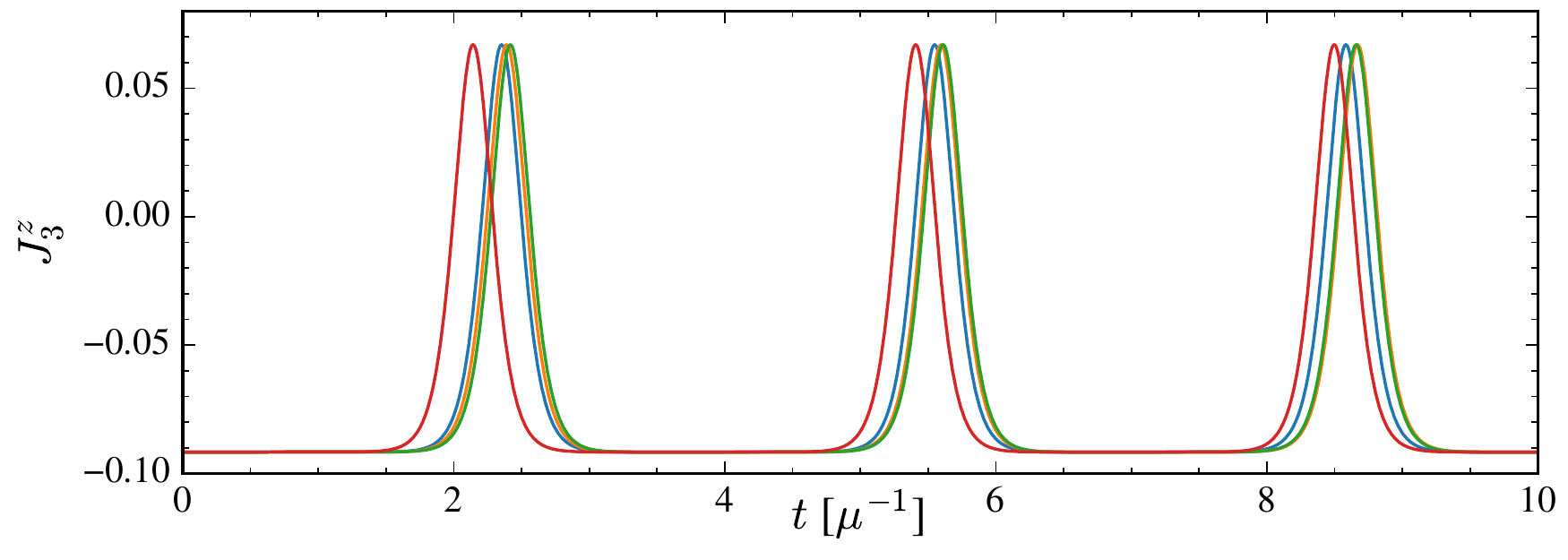}}
		\subfloat[AS fast with bulk velocity]{\includegraphics[width=0.5\textwidth]{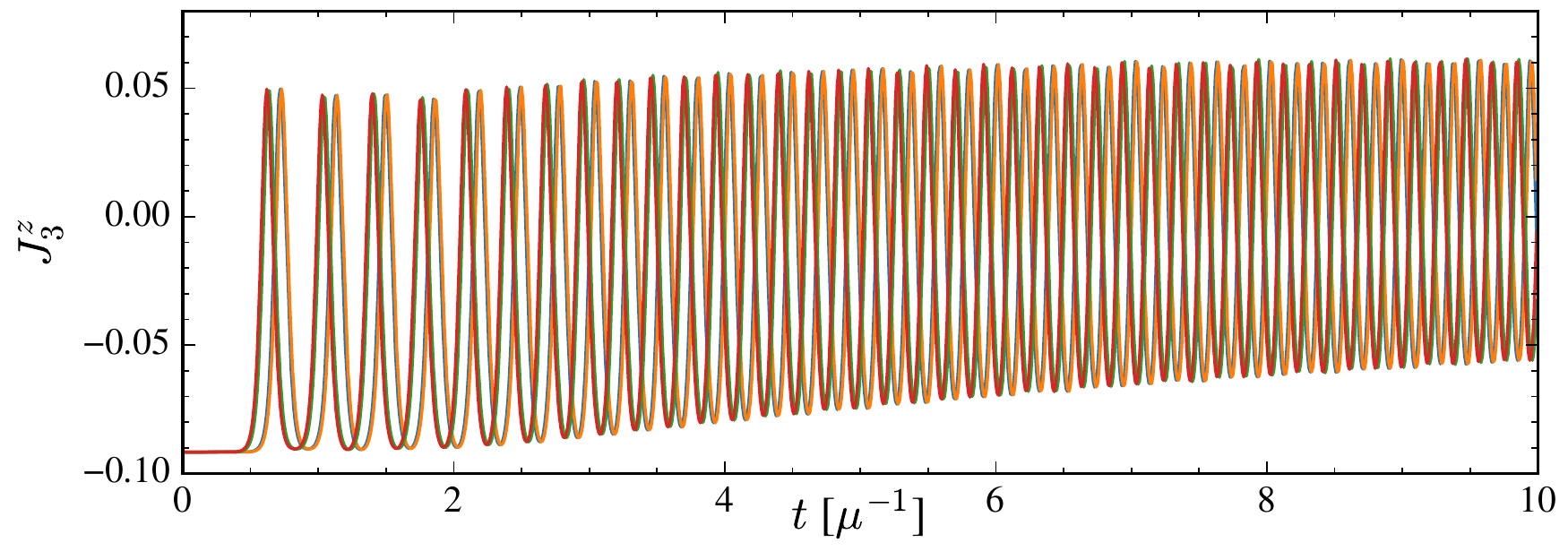}}\\
		\subfloat[MZA slow]{\includegraphics[width=0.5\textwidth]{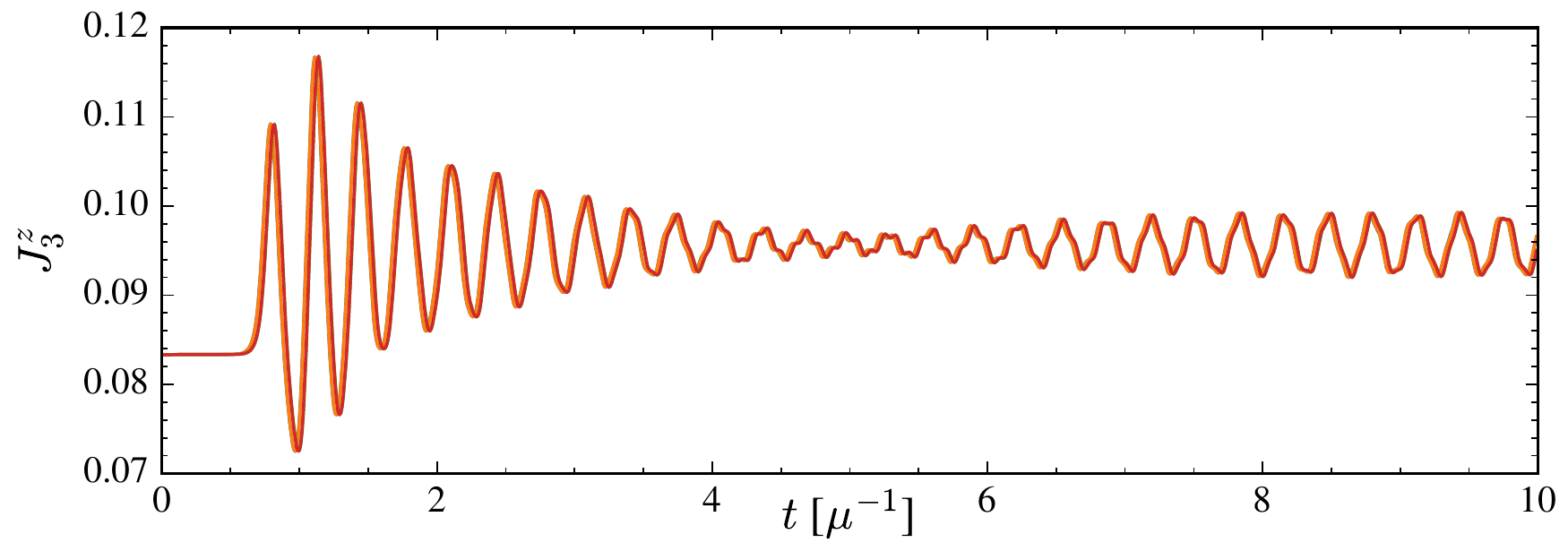}}
		\subfloat[SB fast]{\includegraphics[width=0.5\textwidth]{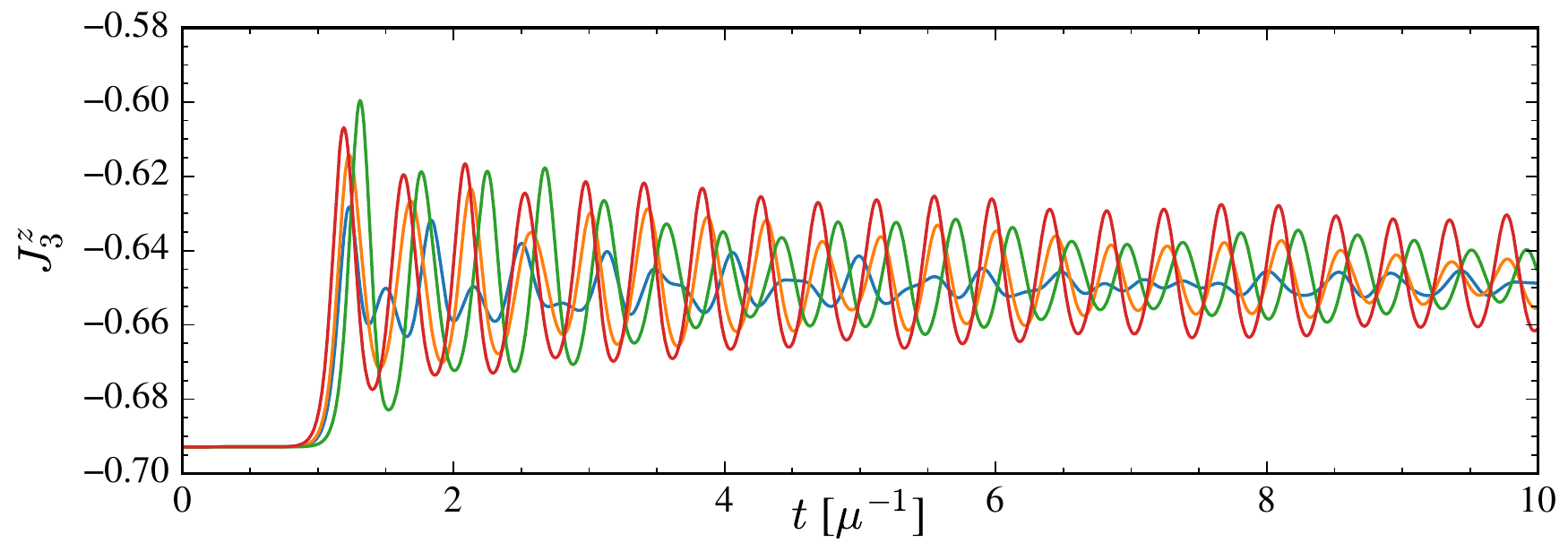}}
	\caption{\label{fig:evolJ} Time evolution of $J^z_3$ for four models. Colors show cases with different initial perturbations in each panel.}
\end{figure*}

The same trend also applies to the components of the polarization currents.
The red curves in Fig.~\ref{fig:evolJ} show $J^z_3$ as a function of time for the pure AS fast mode, AS fast mode with bulk velocity, MZA slow mode, and SB fast mode.
In the single-angle AS fast model, $J^z_3$ evolves in the bipolar manner.
The AS fast model with bulk velocity shows bipolar pattern in the first several cycles but gradually deviates from that pattern.
Both MZA slow and SB fast modes cascade into aperiodic patterns due to the more complicated distribution of their unstable eigenmodes.

\subsection{Dependence on the initial perturbation}
For the mean-field EOM with negligible $\theta_V$, an initial perturbation away from $\hat{\mathbf e}_3$ is needed in order to trigger the collective flavor oscillations.
For cases where the initial perturbations are generated randomly, we check whether the conclusion derived above may sensitively depend on the initial condition. 
Moreover, it may not be guaranteed that these cases will evolve in a way that their evolution paths remain close to each other, which is related to the chaoticity of a non-linear system \cite{hansen2014chaotic}.

Motivated by these considerations, for each model, we take another three randomly generated perturbations, i.e., with different seed $\epsilon_{\rm pert}$.
Their evolutions are shown with different colors in Fig.~\ref{fig:evolJ}.
As expected, whether or not a system behaves as bipolar (AS fast mode) or non-bipolar (the other three cases), is not related to the randomness of the initial perturbation, which is consistent with our arguments based on the geometric symmetry.

For aperiodic models without simple geometric symmetry, one may wonder whether the kinematic decoherence and relaxation can result in very different flavor evolution history for systems with slightly different initial conditions.
Interestingly, we find that the outcomes in the AS fast mode with bulk velocity and MZA slow mode are highly deterministic and repetitive. 
Taking a slightly different initial condition only leads to an overall time off-set for these two modes.
On the other hand, the SB fast model shows significant differences of flavor evolution in nonlinear regime when taking different initial conditions.
It seems to suggest that this case exhibits chaotic feature, which are not present in the other models.
In addition, this dependence on the initial conditions could be linked to the many-body decoherence \cite{xiong2022many,lacroix2022role}.
Since in the many-body description there is always uncertainty to some extent for the polarization vectors in flavor space, the outcome for neutrino flavor evolution may need to be given as the average over the whole ensemble of all possible mean-field evolution trajectories.
Conclusion regarding the chaoticity of these systems requires further dedicated studies, which we defer to the future.

\section{Discussion and conclusions}
\label{sec:conclusions}
We have provided a picture to understand the periodic bipolar flavor evolution for collective neutrino oscillations, which relates to a geometric symmetry of the unstable eigenmode on the two-flavor Bloch sphere.
This symmetry requires a one dimensional circular distribution of the unstable eigenmode in linear regime so that the shape can be maintained on the Bloch spherical surface when the system transitions to the non-linear regime.
An additional constraint on the position of the circular unstable mode is that it should either be \blue{overlapping with the origin on the complex plane, or centered at the origin in the absence of vacuum and matter terms}.
When both requirements are satisfied, the EOMs that govern the growth of the unstable mode in the linear and nonlinear regimes take a similar form up to a scaling factor, which accounts for the spherical bending on the Bloch sphere for the nonlinear case.
Given that, all polarization vectors can synchronously move on the Bloch sphere while retaining their initial circular distribution in a co-rotating frame.
We have derived the analytical solutions to the bipolar motion based on this geometric symmetry.

On the other hand, we have found that this geometric symmetry shared by single-angle slow mode, single-energy fast mode, and four-beam coplanar model is very unusual and strongly depends on the assumptions made for the spectral and angular distributions of the neutrino gas.
We show with numerical examples that other modes such as MZA slow mode and SB fast mode do not share the same bipolar feature but exhibit behavior of kinematic decoherence in flavor evolution because their eigenmodes do not follow a circular distribution on the Bloch sphere.
We have also examined the dependence of our result on the initial random perturbations and found that the connection described above remains. 
Based on this finding, one may expect that for systems that impose less symmetry in the neutrino spectral and angular distributions, e.g., neutrino flux and matter flux not flowing in the same direction as well as the possible existence of quadruple moment in neutrino angular distribution, the bipolar flavor evolution is less likely to occur even for a homogeneous model.

\blue{
In the absence of bulk velocity of matter, the property that the Hamiltonian is linear in the parameter $u$ is the same as that guarantees the conservation of the Gaudin invariants and exact integrability, as identified in Ref.~\cite{fiorillo2023slow}.
Note that the bipolar motion can break down in the presence of bulk velocity of matter.
We further caution that our picture is not equivalent to the pendulum formalism \cite{padilla2022neutrino,fiorillo2023slow} in all aspects.
For example, we mainly focus on the nutation that starts from near the pure flavor state in the linear regime, while the pendulum picture provides a broader description of both nutation and precession.
On the other hand, our picture is applicable to more general spectral and angular distributions, while the pendulum picture focuses on more specific unstable eigenmodes that exhibit the bipolar flavor evolution.}

For a configuration with two or more co-existing unstable modes, the above picture may fail even if each mode matches the above geometric symmetry and can undergo bipolar motion individually.
With the superposition of more modes, the choice of the co-rotating frame becomes ambiguous with different precession frequencies $\Omega_r$, linear growth rates $\Omega_i$, and dropping-down directions for $\mathsf O$.
The interference between each mode may make the multiplicative relation between Eq.~\eqref{eq:bipolar_cp_rr} and Eq.~\eqref{eq:bipolar_pv_rr} no longer valid.

In more general situations where unstable modes are allowed to develop inhomogeneously as $\mathsf P_\perp = \mathsf Q\, e^{-i(\Omega t-\vec K\cdot \vec r)}$, periodic bipolar motion can hardly exist since the symmetry is not fulfilled in a way similar to the AS fast mode with non-zero bulk velocity.
Even in the absence of vacuum term, Eq.~\eqref{eq:pvw_vdrho_P} yields $\partial_\rho \mathbf J^\rho = \lambda^t \hat{\mathbf e}_3 \times \mathbf J^t$ so that $\mathsf{J}^t_\perp$ is not necessarily negligible.
The circular unstable mode does not go through the origin of the complex plane.
Moreover, the extents of deviation also depend on the $\vec K$ mode that dominates the growth in linear regime.
Note that unstable modes can arise in a continuous range of $\vec K$ instead of just $\vec K$=0, which makes inhomogeneous flavor evolution very different from the bipolar behavior.

In addition, the symmetry governing the bipolar motion relies on geometry of the Bloch sphere in a two-flavor oscillation system.
A three-flavor system has more complicated adjoint representation, which also enables more than one mode to develop by distinguishing $\nu_\mu$ and $\nu_\tau$ sectors in spite of the assumed homogeneity.
The growth of the circular unstable eigenmode in, e.g., $\nu_e$--$\nu_\mu$ subspace may be interfered by the unstable mode growing in $\nu_e$--$\nu_\tau$ subspace in non-linear regime, which deforms the periodic bipolar motion.
Similarly, the dynamic decoherence brought by incoherent collisions violates the conservation of the length of neutrino polarization vectors so that each of them does not move on a Bloch sphere of fixed size.
Many further studies are needed for collective neutrino oscillations in all those more general situations.

\appendix

\section{Circular distribution in conformal mapping}
\label{sec:circular_unstable}
Consider a complex conformal mapping function of a real variable $u$ in the form
\begin{equation}
    \mathsf{Q}(u) = \frac{\mathsf{A}+\mathsf{B}h(u)}{\mathsf{C}+\mathsf{D}h(u)},
    \label{eq:Appendix_Qu}
\end{equation}
where $\mathsf{A}$, $\mathsf{B}$, $\mathsf{C}$, and $\mathsf{D}$ are arbitrary complex numbers and $h(u)$ is an arbitrary real function.
\textit{Its distribution is either a straight line or an arc of circle on the complex plane}.
Notice that here $\mathsf{B}$ is in sans serif and indicates an arbitrary complex number (not necessarily related to $\mathbf{B}$ in $\mathbf{H}_\text{vac}$).
Clearly, the distribution of $\mathsf{Q}(u)$ is a line for $\mathsf D=0$.
More generally, when $\mathsf D\neq 0$ but $\mathrm{Im}(\mathsf{C}\,\mathsf{D}^*)=0$, $\mathsf{C}$ and $\mathsf{D}$ have the same argument, and
\begin{equation}
    \mathsf{D}\, \mathsf{Q}(u) =
    \frac{\mathsf{A}-|\mathsf{C}/\mathsf{D}|\mathsf{B}}{|\mathsf{C}/\mathsf{D}|+h(u)} + \mathsf{B}
\end{equation}
represents a line that goes through the point of $\mathsf{B}$ in the direction of $\mathsf{A}-|\mathsf{C}/\mathsf{D}|\mathsf{B}$.

In all the other cases, the distribution of $\mathsf{Q}(u)$ is circular, which can be seen by rewriting Eq.~\eqref{eq:Appendix_Qu} as
\begin{equation}
    \mathsf{Q}(u) = \mathsf O + \mathsf{R} \frac{\mathsf{C}^*+\mathsf{D}^*h(u)}{\mathsf{C}+\mathsf{D}h(u)}.
    \label{eq:Appendix_B3}
\end{equation}
In the above equation,
\begin{align}
    \mathsf{R} & = \frac{\mathsf{A}\, \mathsf{D} - \mathsf{B}\, \mathsf{C}}{2i\,\mathrm{Im}(\mathsf{C}^* \mathsf{D})}, \\
    \mathsf O & = \frac{\mathsf{A}\, \mathsf{D}^* - \mathsf{B}\, \mathsf{C}^*}{2i\,\mathrm{Im}(\mathsf{C}\, \mathsf{D}^*)}. 
    \label{eq:Appendix_B5}
\end{align}
As $|\mathsf{C}^*+\mathsf{D}^*h(u)|/|\mathsf{C}+\mathsf{D}h(u)| = 1$, Eq.~\eqref{eq:Appendix_B3} represents a circle of radius $|\mathsf{R}|$ centered at
the point of $\mathsf O$.
In particular, the circle goes through the origin for $|\mathsf O|=|\mathsf{R}|$, which implies $\mathrm{Re}(\mathsf{A}^*\mathsf{B}\,\mathsf{C}\,\mathsf{D}^*) = \mathrm{Re}(\mathsf{A}\,\mathsf{B}^*\mathsf{C}\,\mathsf{D}^*)$ or equivalently
\begin{equation}
    \mathrm{Im}(\mathsf{A}\,\mathsf{B}^*)=0.
\end{equation}

\begin{acknowledgments}
Z.~X. acknowledges support of the European Research Council (ERC) under the European Union’s Horizon 2020 research and innovation programme (ERC Advanced Grant KILONOVA No.~885281).
M.-R.~W. acknowledges supports from the National Science and Technology Council, Taiwan under Grant No.~110-2112-M-001-050 and 111-2628-M-001-003-MY4, the Academia Sinica under Project No.~AS-CDA-109-M11, and Physics Division, National Center for Theoretical Sciences, Taiwan.
Y.-Z.~Q. acknowledges support from the US Department of Energy under grant DE-FG02-87ER40328.
We acknowledge the use of the software \texttt{Matplotlib} \cite{matplotlib}.
\end{acknowledgments}

\bibliographystyle{apsrev4-2}
\bibliography{references.bib}

\end{document}